\documentclass[a4,12pt]{article}
\usepackage{subcaption}
\usepackage{multirow}
\usepackage{adjustbox}
\usepackage{caption}
\usepackage{graphicx}
\usepackage{amsmath}
\usepackage{amsfonts}
\usepackage{amssymb}
\usepackage{cite}
\usepackage{hyperref}
\usepackage[square,numbers]{natbib}
\usepackage{mathtools}
\usepackage{float}
\usepackage{authblk}

\begin{document}
\title{Effects of  variations  of SUSY breaking scale on neutrino parameters at low  energy scale under radiative corrections}
\date{}
\author[1]{Kh. Helensana Devi \footnote{helensanakhumanthem2@manipuruniv.ac.in}}
\author[1]{K. Sashikanta Singh \footnote{ksm1skynet@gmail.com}}
\author[1,2]{N.Nimai Singh \footnote{nimai03@yahoo.com} }
\affil[1]{\small{{Department of Physics, Manipur University, Imphal-795003, India} }}
\affil[2]{Research Institute of Science and Technology, Imphal-795003, India }

\maketitle \thispagestyle{empty}
\begin{abstract}
{The paper addresses the effects of the variations of the SUSY breaking scale $m_s$ in the range (2-14) TeV on the three neutrino masses and mixings, in running the renormalization group equations (RGEs) for different input values of high energy seesaw scale $M_R$, in both normal and inverted hierarchical neutrino mass models. The present investigation is a continuation of the earlier works based on the variation of $m_s$ scale. Two approaches are adopted one after another - bottom-up approach for running gauge and Yukawa couplings from low to high energy scale, followed by the top-down approach from high to low energy scale for running neutrino parameters defined at high energy scale, along with gauge and Yukawa couplings. A self-complementarity relation among three mixing angles is also employed in the analysis. Significant effect due to radiative corrections on neutrino parameters with the variation of SUSY breaking scale $m_s$, is observed.}
\end{abstract}
{

                      \bf{Keywords}}: Self-complementarity relation, supersymmetry breaking scale, seesaw scale, RGEs, NH case, IH case, SM, MSSM.
\titlepage
\bigskip
\pagebreak
\section{Introduction} 
\hspace{0.25in} Neutrino physics has registered significant progress in recent years with the measurement of non-zero $\theta_{13}$ \cite{a1,a2,a3} and Dirac CP phase \cite{b1, b2}, thus indicating a possibility for a sizable CP violation in neutrino sector. Neutrino oscillations \cite{c, c1,c2} have been well studied with the precise measurements of neutrino mass parameters and mixing angles. But till date there are still some unsettled questions in neutrino physics such as the correct mass hierarchical order whether normal or inverted, absolute neutrino mass scale, nature of neutrino whether Dirac or Majorana type, the exact scale of seesaw mechanism, the supersymmetric breaking scale  if all it exists, to mention a few. The information related to the absolute neutrino masses, has been continuously updating with the recent PLANCK data on cosmological upper bound \cite{g,h} on the sum of three neutrino masses $\Sigma |m_i| < 0.12$ eV, neutrinoless double beta  decay \cite{j2,j1} results with the upper bound in the overall range $m_{ee}< (0.075-0.350)$ eV and KATRIN \cite{k} result on direct kinematic measurement with the upper bound $m_{\nu_e} < 0.8$ eV. Neutrino mass model if any, is bound to be consistent with these upper bounds on absolute neutrino masses.\\
 
    On theoretical front, the presence of supersymmetry (SUSY) \cite{s1,s2,s3} enables to ensure the stability of hierarchy between the weak and GUT scales with the possible cancellation of quadratic term in radiative corrections to the Higgs boson mass. It is needed to have a precise unification point of three gauge couplings at high GUT scale around 2$\times 10^{16}$ GeV \cite{u1,u2,da}. It can also provide a natural mechanism for understanding the electroweak symmetry breaking (EWSB) \cite{ew1} and Higgs physics. Minimal Supersymmetric Standard Model (MSSM) \cite{ms2} is thus a straightforward extension of the Standard Model (SM) with minimum number of new parameters. All the particles in the same supersymmetric multiplet would have the same mass if the supersymmetry is an exact symmetry. So far there is no clear evidence for the presence of supersymmetric particles in the ongoing Large Hadron Colliders (LHC) and LHC has almost reached its maximum energy of about 14 TeV \cite{lhc1,lhc2}. While the existence of supersymmetric particles has been continuously ruling out in LHC, the supersymmetric breaking scale ($m_s$) still remains as an unknown parameter. There are speculations that SUSY particles may have a wide spectrum and are not confined to a single energy scale. For simplicity, one can assume a single scale \cite{ss1,ss2} for all supersymmetric particles and this kind of assumption is valid as long as the $m_z$ or $m_t<<m_s$ \cite{st1,st2}. We assume that the $m_s$ scale may lie somewhere in between 2 TeV and 14 TeV within the scope of LHC. The effects of the variations of SUSY breaking scale on the unification of gauge couplings and also Yukawa couplings in MSSM and SUSY GUT models, have already been addressed using two-loop RGEs for gauge and Yukawa couplings within the minimal supersymmetric SU(5) model\cite{da}, while ignoring for simplicity the threshold effects of the heavy particles, which could be as large as a few percentages. It has already been reported that for gauge couplings, the unification point increases with the increase in the SUSY breaking scale, but for Yukawa couplings the unification points decrease with the increase in SUSY  breaking in the reverse order compared to the gauge couplings \cite{da}. This finding has certain implications in other important issues such as running of the renormalization group equations (RGEs) \cite{rg2, beta, dac} for neutrino masses and mixings from high energy seesaw scale to low energy electroweak scale. In this direction a preliminary analysis with normal hierarchical model has already been carried out on the stability of neutrino parameters and self-complementarity relation \cite{dac} with varying SUSY breaking scale $m_s$.\\
  
   The present investigation is a continuation of our previous work on neutrino masses and mixings with varying SUSY breaking scale in the running of RGEs \cite{da,dac}. We shall address both normal hierarchical and inverted hierarchical neutrino mass models, in both approaches - in the first place, bottom-up approach for running gauge and Yukawa couplings from low to high energy scale, and in the second place, top-down approach for running neutrino parameters defined at high energy scale, along with gauge and Yukawa couplings, from high to low energy scale. A self-complementarity relation among three mixing angles, $\displaystyle \theta_{23}\approx \theta_{12} + \theta_{13}$ is also employed in the analysis. \\
   
   The paper is organised as follows. In section 2 we give a brief discussion of gauge and Yukawa coulings RGEs mainly on bottom-up and top-down runnings. In section 3, we present the numerical analysis and results. Summary and discussion are presented in section 4. We give relevant RGEs for gauge, Yukawa and quartic Higgs couplings in two-loops fo both the SM and MSSM in Appendix A and RGEs of neutrino parameters in Appendix B.
\section{Renormalisation Group Equations (RGEs)}
We study the radiative corrections to neutrino parameters using Renormalisation Group Equations (RGEs) \cite{da,rg1, beta} with and without SUSY in two different steps using the low energy observational input values, bottom-up running from low to high energy scale for gauge and Yukawa couplings, and top-down running from high to low energy scale for neutrino mass parameters and mixing angles, along with gauge and Yukawa couplings which are already evaluated at high energy scale $M_R$.
\subsection{Bottom-up Running}

 In the bottom-up running of the RGEs, we divide it into three regions, $m_z < \mu < m_t$, $m_t < \mu <m_s$, $m_s< \mu <M_R$. We use recent experimental data \cite{d,c2} as initial input values at low energy scale, given in Table \ref{tab:1z}. 
 
\begin{table}[H]
\begin{adjustbox}{width=0.99\textwidth}
\centering
\begin{tabular}{lll}
\hline
Mass(GeV)& Coupling constants & Weinberg mixing angle \\ 
\hline
$m_z(m_z)$= 91.1876& $\alpha_{em}^{-1}(m_z)$ =127.952 $\pm$ 0.009 & $\sin^2\theta_w(m_z)$ = 0.23121 $\pm 0.00017$\\
 $m_t(m_t)$= 172.76 & $\alpha_{s}(m_z)$ =0.1179$\pm$ 0.009\\
 $m_b(m_b)$= 4.18 &&\\
 $m_{\tau}(m_{\tau})$=1.77&&\\
\hline 
\end{tabular} 
\end{adjustbox}
\caption{ \footnotesize{Latest experimental data for fermion masses, gauge coupling constants and Weinberg mixing angle.}}
\label{tab:1z}
\end{table}
 
  The values of gauge couplings, $\alpha_{2}$  for $SU(2)_{L}$ and $\alpha_{1}$ for $U(1)_{Y}$, are calculated by using  $\displaystyle \sin^2 \theta_{w}(m_{z}) = \frac{\alpha_{em}(m_z)}{\alpha_2(m_z)}$ and matching condition, 
\begin{equation}
\frac{1}{\alpha_{em}(m_z)} =\frac{5}{3}\frac{1}{\alpha_1(m_z)} + \frac{1}{\alpha_2(m_z)}.
\label{b}
\end{equation}
We can also express the gauge couplings $\alpha_i$'s \cite{da} in terms of normalized couplings $g_i$'s as $g_i = \sqrt{4\pi \alpha_i}$, where $i=1, 2, 3$ denote electromagnetic, weak and strong couplings respectively. RGEs at one-loop level \cite{bb1} is used for evolution of the three gauge coupling constants from $m_z$ scale to $m_t$ scale, as given below:
\begin{equation}
\frac{1}{\alpha_i(\mu)} = \frac{1}{\alpha_{i}(m_z)} - \frac{b
_i}{2\pi}ln\frac{\mu}{m_z},
\label{d}
\end{equation}
where $m_z\leq \mu \leq m_t$  and $b_i$ = (5.30, -0.50, -4.00) for non-SUSY case. For fermion masses to define at $m_t$ scale, we use QED-QCD rescaling factor $\eta$  \cite{qcd}, $\displaystyle m_b(m_t) = \frac{m_b(m_b)}{\eta_b}$ and $\displaystyle  m_{\tau}(m_t) = \frac{m_{\tau}(m_\tau)}{\eta_{\tau}}$, where $\displaystyle  \eta_{b} = 1.53$ and $\eta_{\tau} = 1.015$. We then convert them to Yukawa couplings, $\displaystyle  h_t(m_t) = \frac{m_t(m_t) }{v_0}$, $\displaystyle  h_b(m_t) = \frac{m_b(m_b)}{v_0 \eta_b}$, and $\displaystyle  h_\tau(m_\tau) = \frac{m_\tau(m_\tau)}{v_0\eta_\tau}$, where $v_0$ = 174 GeV is the vacuum expectation of SM Higgs field. The calculated numerical values for fermion masses, Yukawa and gauge couplings at $m_t$ scale are given in Table \ref{tab:1b}.

\begin{table}[H]
\centering
\begin{tabular}{ccc} 
\hline
Fermions masses & Yukawa Couplings & Gauge Couplings \\
\hline
$m_t(m_t)$ = 172.76 GeV& $h_t(m_t)$ = 0.9928&$g_1(m_t)$ =  0.4635\\
$m_b(m_t)$ = 2.73 GeV & $h_b(m_t)$ = 0.0157&$g_2(m_t)$ = 0.6511\\
$m_{\tau}(m_t)$ = 1.75 GeV  & $h_{\tau}(m_t)$ = 0.0100&$g_3(m_t)$ = 1.1890\\
\hline
\end{tabular} 
\caption{\footnotesize{ Numerical values for fermion masses, Yukawa  and gauge couplings at $m_t$ scale.}}
\label{tab:1b}
\end{table}

We study the effect of variation of SUSY breaking scale ($m_s$) on gauge and Yukawa couplings for running from $m_t$ to the $M_R$ scale using RGEs which are given in Appendix A. At $m_s$ scale, the following matching conditions are applied at the transition point from SM ($m_t<\mu<m_s$) to MSSM ($m_s<\mu<M_R$) as

\begin{equation}
\left.
\begin{array}{l}
g_i(SUSY) = g_i(SM) \\
    h_t(SUSY) = \frac{h_t(SM)}{\sin\beta}= h_t(SM) \times \frac{\sqrt{1+ \tan^2\beta}}{\tan\beta}\\
h_b(SUSY) = \frac{h_b(SM)}{\cos\beta}= h_b(SM) \times \sqrt{1+ \tan^2\beta}\\
h_\tau(SUSY) = \frac{h_\tau(SM)}{\cos\beta}= h_\tau(SM) \times \sqrt{1+ \tan^2\beta} 
\end{array}
\right\}
\end{equation}

The output for Yukawa and gauge couplings at $M_R$ scale, are given in Table \ref{tab:1c} for $M_R = 10^{13}$ GeV, Table \ref{tab:1d} for $M_R = 10^{14}$ GeV, Table \ref{tab:1e} for $M_R = 10^{15}$ GeV and Table \ref{tab:1f} for $M_R = 10^{16}$ GeV respectively, for common value of $\tan \beta$ = 40. These values are needed for the next top - down running as input values at high energy scale.

\begin{table}[H]
\centering
\begin{tabular}{cccc|ccc}
\hline 
$m_s$(TeV) &$h_t$ &$h_b$ &$h_\tau$ &$g_1$ &$g_2$& $g_3$\\
\hline
2&   0.6509  &0.3139  &0.3721 & 0.6086 &0.6914 &0.7743 \\ 
4&   0.6318  &0.3102  &0.3741 &0.6043  & 0.6861 & 0.7700 \\ 
6 &   0.6236  &0.3086  &0.3750   &0.6023&   0.6835 &0.7679 \\
8&      0.6161 &  0.3071   &  0.3760 &0.6002 &     0.6810&   0.7658  \\ 
10 &  0.6126  & 0.3063   & 0.3765   &0.5992  & 0.6797 &0.7648  \\  
12 &    0.6093   & 0.3056 &  0.3770& 0.5981 &0.6785    &0.7638  \\
14 & 0.6061  & 0.3050 &   0.3775  &  0.5971  &  0.6772 &0.7628 \\
\hline 
\end{tabular} 
\caption{ \footnotesize{Values of Yukawa and gauge couplings evaluated at  $t_R = ln(10^{13}$ GeV) = 29.93 for $\tan\beta$ = 40}, for different choices of $m_s$ scale. }
\label{tab:1c}
\end{table}
\begin{table}[H]
\centering
\begin{tabular}{cccc|ccc}
\hline 
$m_s$(TeV) &$h_t$ &$h_b$ &$h_\tau$ &$g_1$ &$g_2$& $g_3$\\
\hline
2&0.6289&  0.2977&   0.3650&   0.6316&   0.6982& 0.7554  \\ 
4&   0.6104&   0.2946&   0.3674&   0.6268&    0.6915& 0.7514 \\ 
6 &0.6022&    0.2931&    0.3684&    0.6245&    0.6889& 0.7495  \\ 
8&  0.5947&   0.2917&     0.3695&    0.6222&    0.6863& 0.7475 \\ 
10 &  0.5913& 0.2910&    0.3700&    0.6211&    0.6850& 0.7466 \\  
12 &   0.5879&  0.2904&    0.3705&    0.6199&   0.6837& 0.7456\\
14 & 0.5848&  0.2898&   0.3710&    0.6188&   0.6824& 0.7447  \\
\hline 
\end{tabular} 
\caption{ \footnotesize{Values of Yukawa and gauge couplings  evaluated at $t_R = ln(10^{14}$ GeV) = 32.23 for $\tan\beta$ = 40}, for different choices of $m_s$ scale.}
\label{tab:1d}
\end{table}
\begin{table}[H]
\centering
\begin{tabular}{cccc|ccc}
\hline 
$m_s$(TeV) &$h_t$ &$h_b$ &$h_\tau$ &$g_1$ &$g_2$& $g_3$\\
\hline
2& 0.6084 &   0.2829& 0.3579& 0.6574& 0.7026 & 0.7378 \\ 
4& 0.5891 & 0.2798  &0.3601&  0.6521  &0.6971&  0.7341 \\ 
6 & 0.5809 &  0.2784 &  0.3613&  0.6494 & 0.6944&  0.7323\\
8&   0.5735  &0.2771 &0.3624  &0.6468 &0.6917  &0.7305  \\ 
10 & 0.5701 & 0.2766 & 0.3629 & 0.6455 & 0.6903 & 0.7296 \\  
12 & 0.5668  &0.2760 &0.3635  &0.6443   & 0.6890 &  0.7287  \\
14 &   0.5637 &0.2754& 0.3640 &0.6430 &  0.6877  &0.7278 \\
\hline 
\end{tabular} 
\caption{ \footnotesize{Values of Yukawa and gauge couplings   $t_R = ln(10^{15}$ GeV) = 34.53 for $\tan\beta$ = 40, for different choices of $m_s$ scale.}}
\label{tab:1e}
\end{table}
\begin{table}[H]
\centering
\begin{tabular}{cccc|ccc}
\hline 
$m_s$(TeV) &$h_t$ &$h_b$ &$h_\tau$ &$g_1$ &$g_2$& $g_3$\\
\hline
2 & 0.5854    & 0.2676& 0.3494 & 0.6893 &0.7089   &0.7200\\
4  & 0.5661  &0.2647  &0.3518  &0.6831  &0.7032   &0.7166\\ 
6   &  0.5580 &  0.2634&   0.3529  & 0.6801&   0.7004 &0.7149  \\ 
 8   & 0.5680   & 0.2687&   0.3622 & 0.6770 &   0.6664 &0.7131 \\  
10    & 0.5473 &  0.2618 &  0.3547  &0.6757 &  0.6963 &0.7124\\
12  & 0.5441  &  0.2613   &  0.3553&   0.6742&   0.6949& 0.7116\\
14  & 0.5410 &   0.2608   &0.3559 &  0.6727   & 0.6936 &0.7107\\ 
\hline 
\end{tabular} 
\caption{ \footnotesize{Values of Yukawa and gauge couplings evaluated at  $t_R = ln(10^{16}$ GeV) = 36.84 for $\tan\beta$ = 40, for different choices of $m_s$ scale.}}
\label{tab:1f}
\end{table}
\subsection{Top-down Running}
 In this running, we use the values of Yukawa and guage couplings which are found at $M_R$ scale as initial inputs. We give the sum of three neutrino masses in the range, $\Sigma|m_i| \approx $ 0.11 eV for NH case and $\Sigma |m_i| \approx $ 0.10 eV for IH case. Using all the necessary mathematical frameworks, we analyze the radiative nature of neutrino parameters like neutrino masses, mixings, CP phases, using top-down approach  with the variations of $m_s$ scale at different $M_R$ scale, using respective RGEs which are given in Appendix B. The input sets are given in Table \ref{tab:2} and Table \ref{tab:3}.
 
\begin{table}[H]
\centering
\begin{tabular}{lllll}
\hline 
Input &  \multicolumn{4}{c}{Seesaw scale ( $\tan \beta= 40^0$)} \\
parameters &$10^{16}$ GeV&$10^{15}$ GeV &$10^{14}$ GeV&$10^{13}$ GeV\\
\hline
$m_1$(eV)&0.0262&0.0258&0.0271& 0.0274 \\ 
$m_2$(eV)& -0.0263&-0.0259& -0.0272& -0.0275 \\ 
$m_3$(eV) & -0.0615&-0.0645& -0.0643& -0.0664\\ 
$|\Sigma m_i|$ &0.114&0.116&0.118&0.121\\
$\theta_{12}/^0$&32.46&33.95& 33.01&32.61\\ 
$\theta_{13}/^0$ &7.39& 7.56&7.64 &7.70\\  
$\psi/^0$ &180&180& 180&180\\
$\delta/^0$&240&  240 &240 &240\\
\hline 
\end{tabular} 
\caption{ \footnotesize{Input set of neutrino parameters at high energy scale $M_R$ for NH case. $\theta_{23}$ is used from SC relation, $ \theta_{23}$ = q$\times (\theta_{13}$ + $\theta_{12}$) with q=1.1. This is common for all cases of $m_s$ scale.}}
\label{tab:2}
\end{table}
\begin{table}[H]
\centering
\begin{tabular}{lllll}
\hline 
Input &  \multicolumn{4}{c}{Seesaw scale ( $\tan \beta= 40^0$
)} \\
parameters &$10^{16}$ GeV&$10^{15}$ GeV &$10^{14}$ GeV&$10^{13}$ GeV\\
\hline
$m_1$(eV)&0.0515&0.0501&0.0511& 0.0523  \\ 
$m_2$(eV)& -0.0516&-0.0502& -0.0512& -0.0524 \\ 
$m_3$(eV) & -0.0025&-0.0021& -0.0022& -0.0025\\
$|\Sigma m_i|$ &0.1056&0.1024&0.1045&0.1072\\ 
$\theta_{12}/^0$&31.94&32.39& 31.99&32.17\\ 
$\theta_{13}/^0$ &8.53& 8.29&8.35 &8.40\\ 
$\psi/^0$ &180&180& 180&180\\
$\delta/^0$&240&  240 &240 &240\\
\hline 
\end{tabular} 
\caption{ \footnotesize{Input set of neutrino parameters at high energy scale $M_R$ for IH case ($m_3\neq 0$). $\theta_{23}$ is used from SC relation, $\theta_{23}$ = q$\times (\theta_{13}$ + $\theta_{12}$) with q=1.1. This is common for all cases of $m_s$ scale.}}
\label{tab:3}
\end{table}

\section{Numerical Analysis  and Results}
\begin{figure}[H]
\begin{subfigure}{0.16\textwidth}
    \includegraphics[scale=0.78]{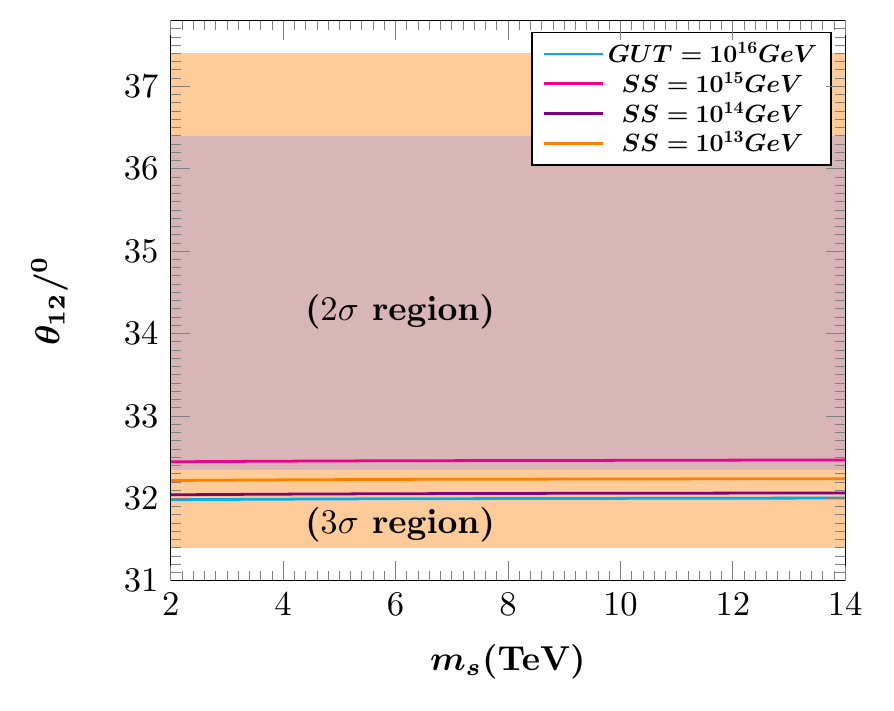}
\end{subfigure}
\begin{subfigure}{0.45\textwidth}
    \includegraphics[scale=0.78]{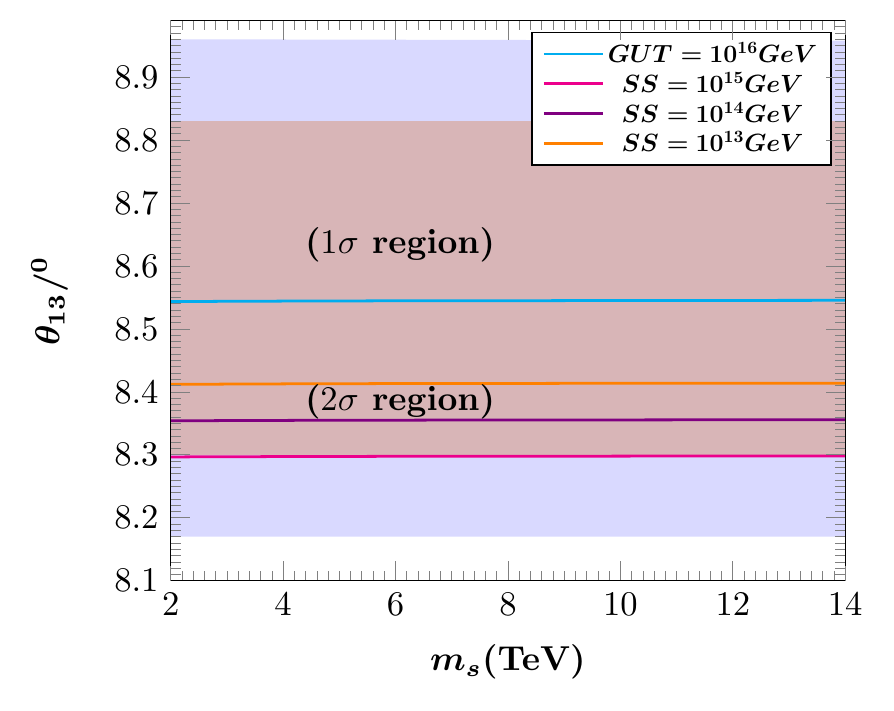}
\end{subfigure}
\begin{subfigure}{.16\textwidth}
    \includegraphics[scale=0.78]{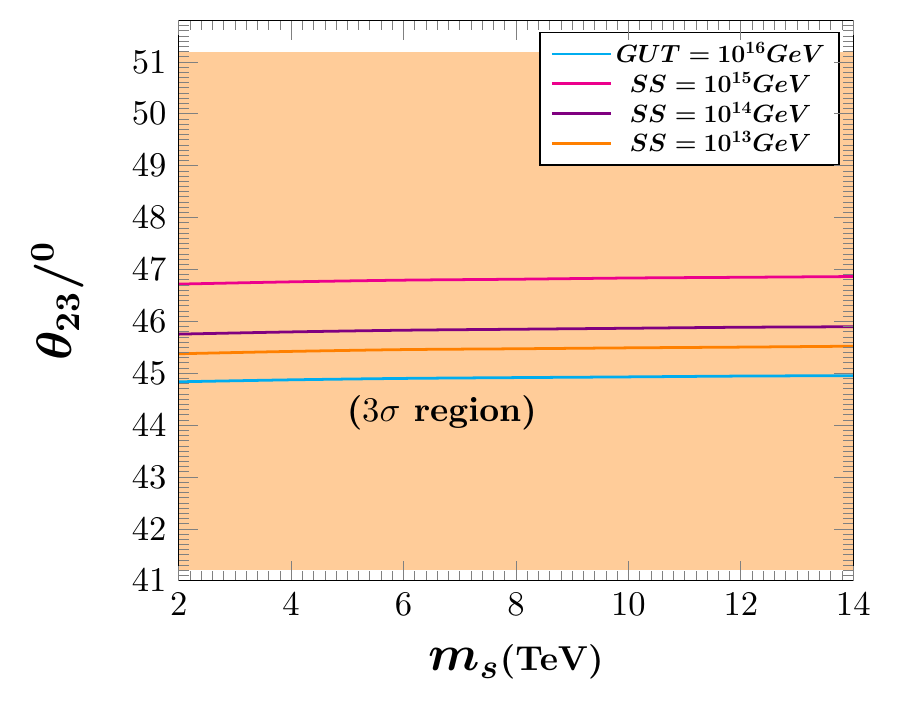}
\end{subfigure}
\begin{subfigure}{0.45\textwidth}
   \includegraphics[scale=0.78]{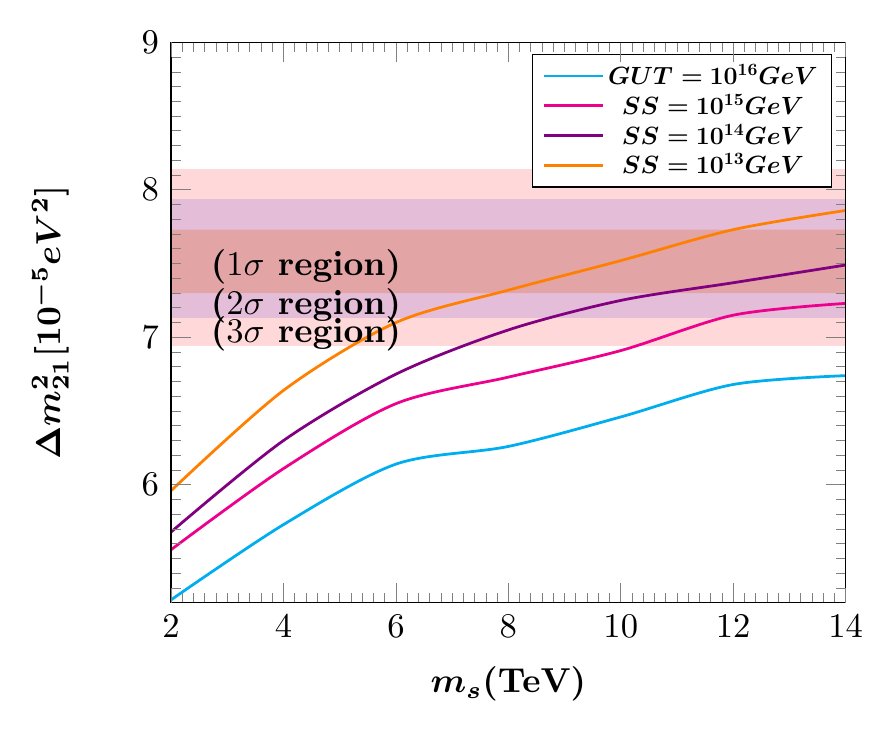}
\end{subfigure}
\begin{subfigure}{0.55\textwidth}
  \includegraphics[scale=0.78]{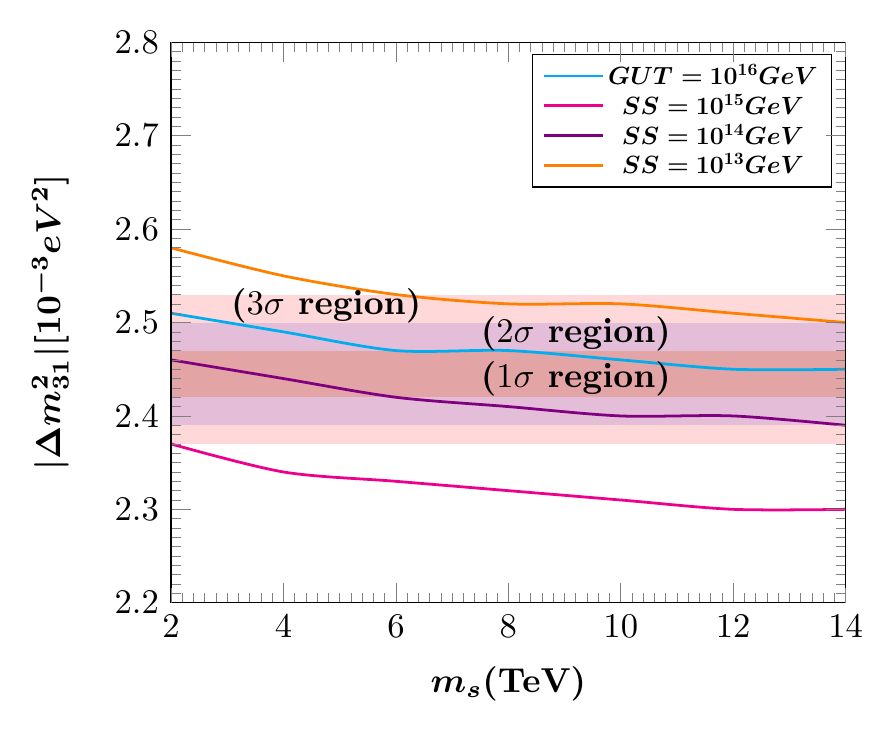}
\end{subfigure}
\begin{subfigure}{0.35\textwidth}
  \includegraphics[scale=0.78]{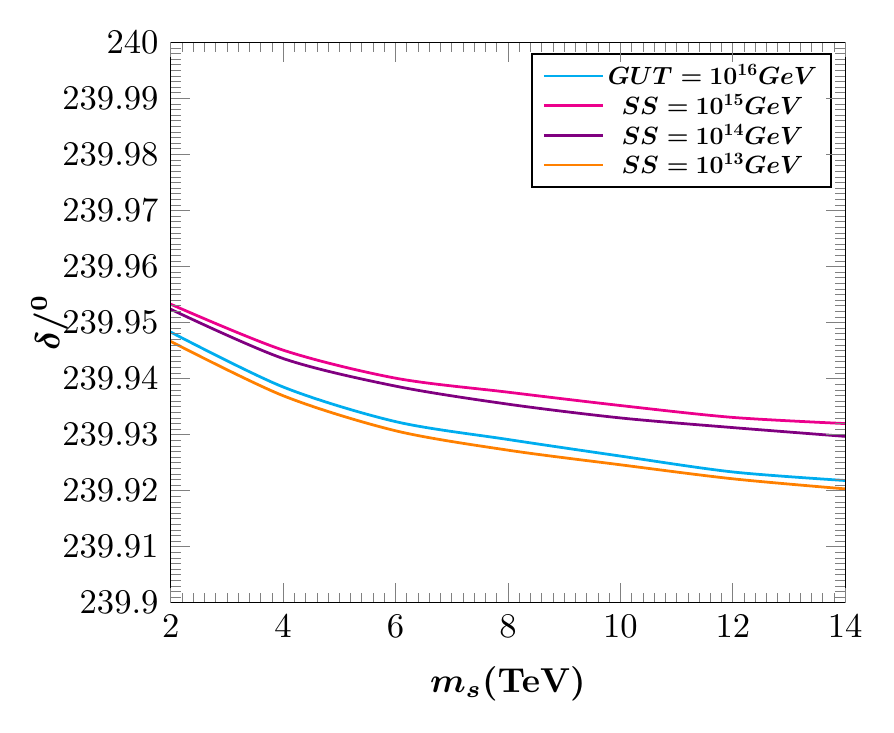}
\end{subfigure}

\caption{\footnotesize{Effects on the low energy output results in $ \theta_{ij}$, $ |\Delta m^{2}_{ij}|$ and $\delta$ with variation of $m_s$ for IH (QD) case at  $\tan \beta =40$. Four different choices of $M_R$ scale are presented.}}
\label{fig:fig1}
\end{figure}

\begin{figure}[H]
\centering
\begin{subfigure}{0.54\textwidth}
    \includegraphics[scale=0.78]{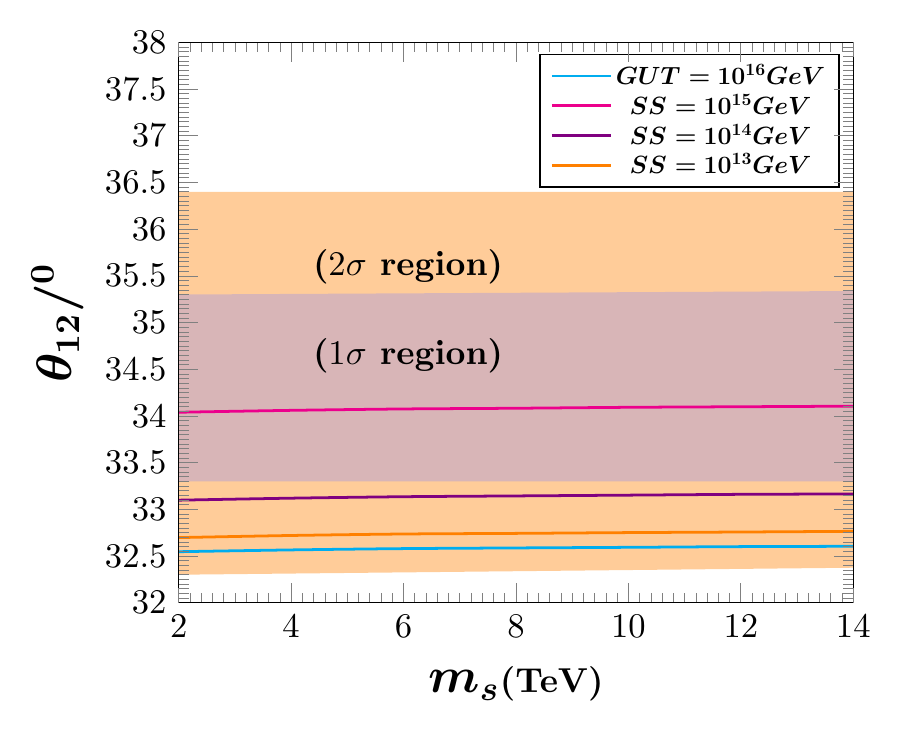}
\end{subfigure}
\begin{subfigure}{0.43\textwidth}
    \includegraphics[scale=0.78]{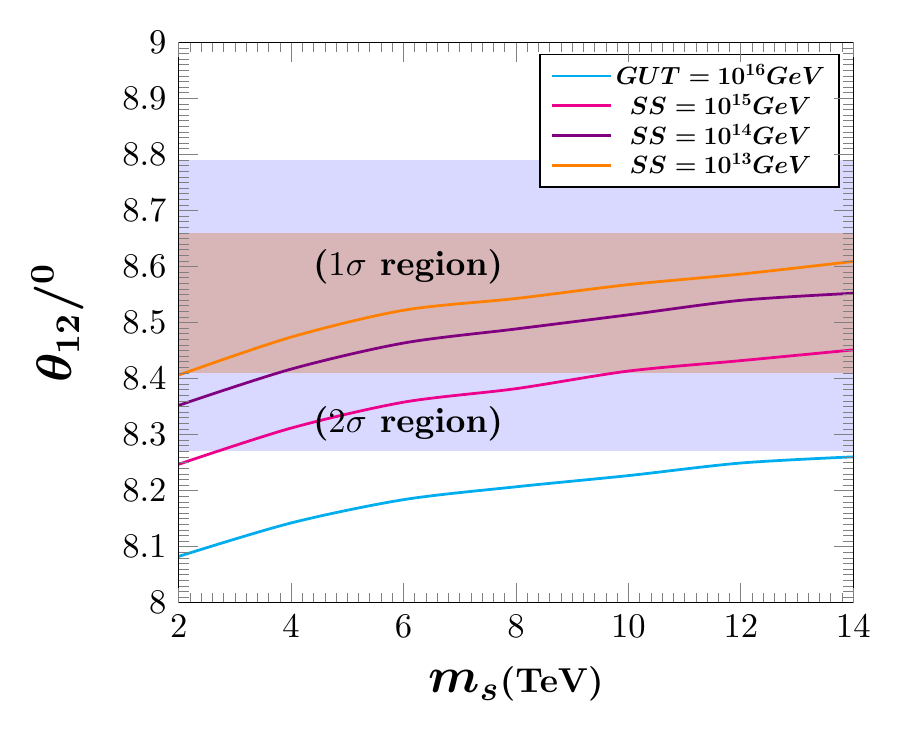}
\end{subfigure}
\begin{subfigure}{.54\textwidth}
    \includegraphics[scale=0.78]{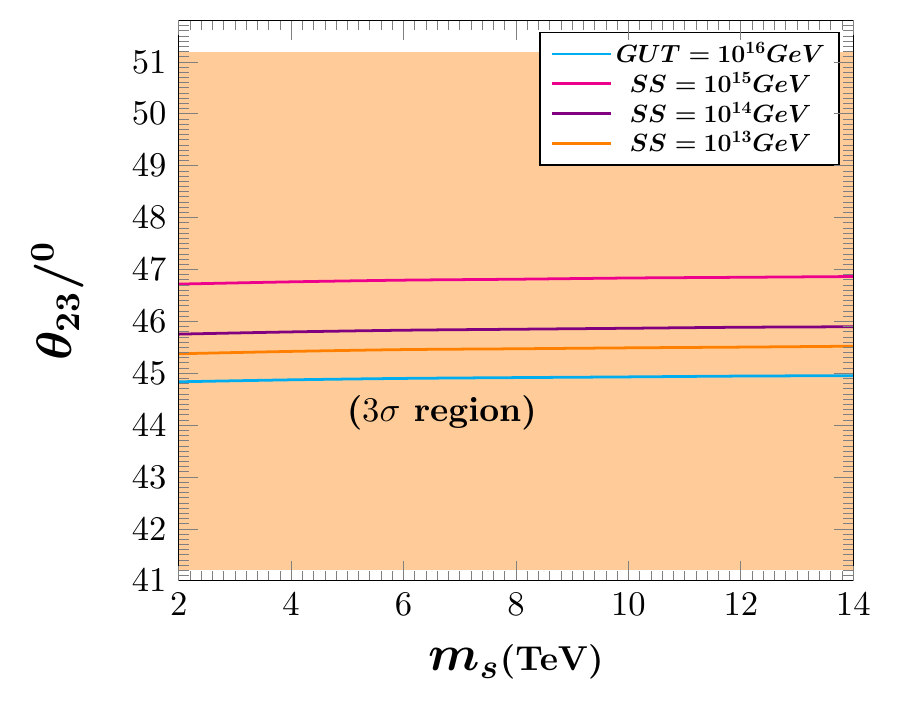}
\end{subfigure}
\begin{subfigure}{0.43\textwidth}
   \includegraphics[scale=0.78]{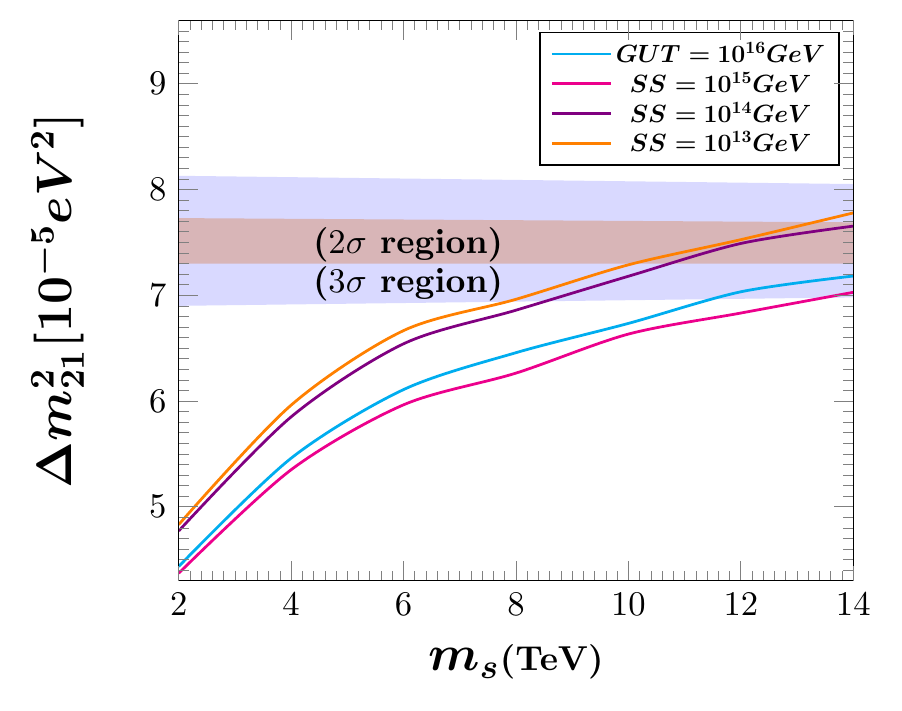}
\end{subfigure}
\begin{subfigure}{0.54\textwidth}
  \includegraphics[scale=0.78]{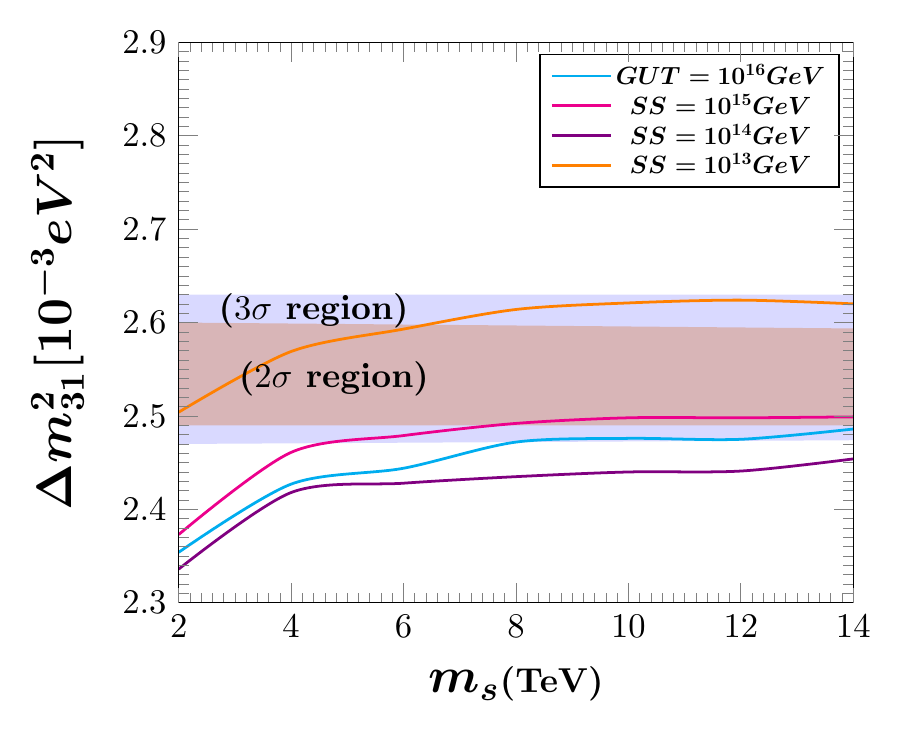}
\end{subfigure}
\begin{subfigure}{0.43\textwidth}
  \includegraphics[scale=0.78]{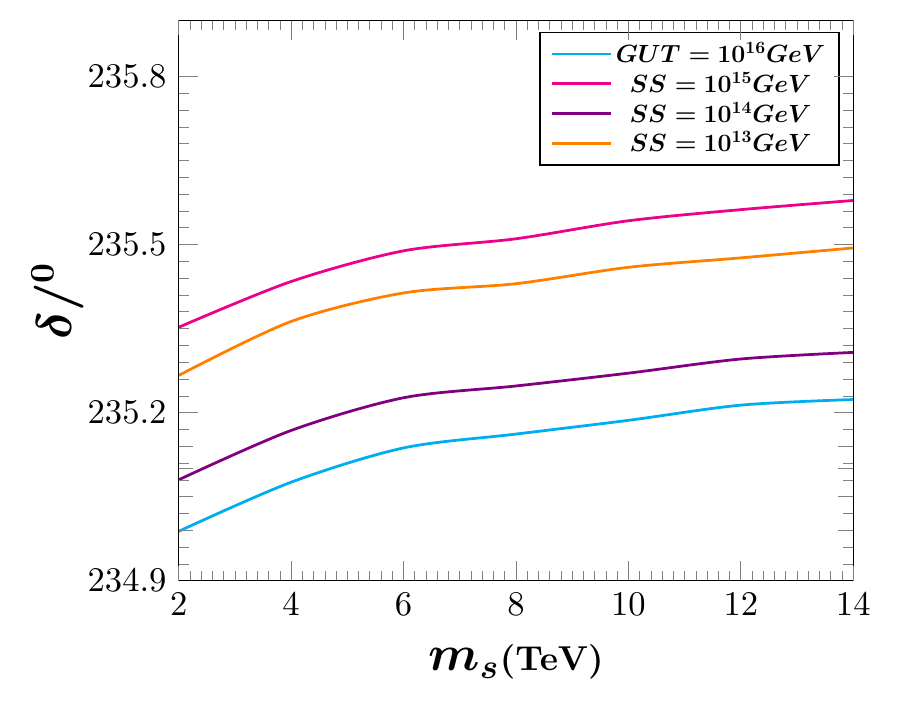}
\end{subfigure}

\caption{\footnotesize{Effects on the low energy output results in $ \theta_{ij}$, $ |\Delta m^{2}_{ij}|$ and $\delta$ with variation of $m_s$ for NH(QD) case at  $\tan \beta =40$. Four different choices of $M_R$ scale are presented.}}
\label{fig:fig2}
\end{figure}

     The effects of the variation of $m_s$ on the outputs of neutrino mass parameters and mixing angles are given in Tables \ref{tab:4} - \ref{tab:7}, along with the graphical representations in Figures \ref{fig:fig2} for normal hierarchical (NH) model; and in Tables \ref{tab:8} - \ref{tab:11} and Figures \ref{fig:fig1} for inverted hierarchical (IH) case. In each case we also present the results for variation of high energy seesaw scale ($10^{13}$ - $10^{16}$) GeV. Similar patterns with the variations of seesaw scale are observed in all the Figures \ref{fig:fig1} and \ref{fig:fig2}. 
     
\begin{table}[H]
\begin{tabular}{cccccccc}
\hline  
$m_s$ scale &$\Delta m^2_{31}$ &$\Delta m^2_{21}$ &$\theta_{23}$&$\theta_{12}$& $\theta_{13}$&$\delta$ &$|\Sigma m_i|$  \\
(TeV) & $(10^{-3}eV^2$)&$(10^{-5}eV^2$)&$(/^0$)&$(/^0$)&$(/^0$)&$(/^0$)&(eV)\\
\hline
2 & 2.504&4.831&45.37 &  32.69 & 8.40&  235.26 &0.095\\ 
4 & 2.569& 5.961& 45.42&  32.72 & 8.47&  235.36 &0.094\\ 
6 & 2.593 & 6.666& 45.45&32.73& 8.52&  235.41&0.094\\ 
8 &2.614 & 6.962& 45.47 &  32.74 & 8.54 & 235.43&0.094\\ 
10 &2.621&7.288&45.48& 32.75 &8.56 & 235.45&0.093\\ 
12 & 2.624& 7.526 &45.50 & 32.75 & 8.58&  235.47&0.093\\ 
14 &2.620&7.779&45.52& 32.76&8.60& 235.49&0.092\\
\hline 
\end{tabular} 
\caption{ \footnotesize{ Effects on the output of neutrino parameters at low energy scale, on varying $m_s$ for NH case ($\tan\beta =40$, $M_R \sim 10^{13}$ GeV).}}
\label{tab:4}
\end{table}

\begin{table}[H]
\begin{tabular}{cccccccc}
\hline  
$m_s$ scale &$\Delta m^2_{31}$ &$\Delta m^2_{21}$ &$\theta_{23}$&$\theta_{12}$& $\theta_{13}$&$\delta$ &$|\Sigma m_i|$     \\
(TeV) & $(10^{-3}eV^2$)&$(10^{-5}eV^2$)&$(/^0$)&$(/^0$)&$(/^0$)&$(/^0$)&(eV)\\
\hline
2 & 2.336&4.771&45.75 & 32.09 & 8.35&  235.08&0.092 \\ 
4 & 2.418&5.850&45.79&  32.11 & 8.41& 235.16 &0.093\\ 
6 & 2.428& 6.541& 45.82&33.13&   8.46&  235.22&0.092\\ 
8 &2.435 & 6.859 & 45.84&  33.14 & 8.48 &  235.24&0.091\\ 
10 &2.440&7.180 &45.86 & 33.15 &8.51 & 235.27&0.091\\ 
12 & 2.441& 7.490 &45.88 &33.16 & 8.53& 235.29&0.090\\ 
14 &2.454&7.654&45.89& 33.16&  8.55&  235.30&0.090\\
\hline 
\end{tabular} 
\caption{ \footnotesize{Effects on the output of neutrino parameters at low energy scale, on varying $m_s$ for NH case ($\tan\beta =40$, $M_R \sim 10^{14}$ GeV).}}
\label{tab:5}
\end{table}

\begin{table}[H]
\begin{tabular}{cccccccc}
\hline  
$m_s$ scale &$\Delta m^2_{31}$ &$\Delta m^2_{21}$ &$\theta_{23}$&$\theta_{12}$& $\theta_{13}$&$\delta$   &$|\Sigma m_i|$  \\
(TeV) & $(10^{-3}eV^2$)&$(10^{-5}eV^2$)&$(/^0$)&$(/^0$)&$(/^0$)&$(/^0$)&(eV)\\
\hline
2 & 2.373&4.371&46.71 & 34.03 & 8.24&  235.35 &0.090\\ 
4 & 2.461&5.350&46.76&  34.05 & 8.31&  235.43& 0.090\\ 
6 & 2.479& 5.964& 46.79&34.07& 8.35&  235.48&0.090\\ 
8 &2.492 & 6.264 & 46.81&  34.08 & 8.38 & 235.51&0.090\\ 
10 &2.498&6.633 &46.83 & 34.09 &8.41 &235.54&0.089\\ 
12 & 2.498& 6.832 &46.84 & 34.09& 8.43&  235.56&0.089\\ 
14 &2.499&7.028&46.86& 34.10& 8.45&  235.57&0.088\\
\hline 
\end{tabular} 
\caption{\footnotesize{ Effects on the output of neutrino parameters at low energy scale, on varying $m_s$ for NH case ($\tan\beta =40$, $M_R \sim 10^{15}$ GeV).}}
\label{tab:6}
\end{table}

\begin{table}[H]
\begin{tabular}{cccccccc}
\hline  
$m_s$ scale &$\Delta m^2_{31}$ &$\Delta m^2_{21}$ &$\theta_{23}$&$\theta_{12}$& $\theta_{13}$&$\delta$  &$|\Sigma m_i|$   \\
(TeV) & $(10^{-3}eV^2$)&$(10^{-5}eV^2$)&$(/^0$)&$(/^0$)&$(/^0$)&$(/^0$)&(eV)\\
\hline
2 & 2.354&4.435&44.83& 32.54 & 8.08& 234.98 &0.093\\ 
4 & 2.427&5.459&44.87&  32.56 & 8.14&  235.07& 0.094\\ 
6 & 2.444& 6.108& 44.89&32.57&   8.18&  235.13&0.094\\ 
8 &2.472 & 6.457 & 44.91&  32.58 & 8.20 & 235.16&0.094\\ 
10 &2.476&6.734 &44.92 & 32.59 &8.22 & 235.18&0.093\\ 
12 & 2.475& 7.033 &44.94 & 32.60 &8.24& 235.21&0.093\\ 
14 &2.486&7.182&44.95& 32.60& 8.26& 235.22&0.093\\
\hline 
\end{tabular} 
\caption{\footnotesize{ Effects on the output of neutrino parameters at low energy scale, on varying $m_s$ for NH case ($\tan\beta =40$, $M_R \sim 10^{16}$ GeV).}}
\label{tab:7}
\end{table}

\begin{table}[H]
\begin{tabular}{cccccccc}
\hline  
$m_s$ scale &$|\Delta m^2_{31}|$ &$\Delta m^2_{21}$ &$\theta_{23}$&$\theta_{12}$& $\theta_{13}$&$\delta$  &$|\Sigma m_i|$   \\
(TeV) & $(10^{-3}eV^2$)&$(10^{-5}eV^2$)&$(/^0$)&$(/^0$)&$(/^0$)&$(/^0$)&(eV)\\
\hline
2 & 2.58&5.96&44.70 & 32.21 & 8.41& 239.94 &0.119\\ 
4 & 2.55&6.64&44.71&  32.22 & 8.41&  239.93 &0.117\\ 
6 & 2.53& 7.10& 44.72&32.22&   8.41&    239.93&0.116\\ 
8 &2.52 & 7.32 & 44.73& 32.23 & 8.41 &  239.92&0.115\\ 
10 &2.52&7.52 &44.73 & 32.23 &8.41 &  239.92&0.114\\ 
12 & 2.51& 7.73 &44.74 & 32.23 & 8.41&   239.92&0.114\\ 
14 &2.50&7.86&44.74& 32.23& 8.41&   239.92&0.113\\
\hline 
\end{tabular} 
\caption{\footnotesize{ Effects on the output of neutrino parameters at low energy scale, on varying $m_s$ for IH case ($\tan\beta =40$, $M_R \sim 10^{13}$ GeV).}}
\label{tab:8}
\end{table}

\begin{table}[H]
\begin{tabular}{cccccccc}
\hline  
$m_s$ scale &$|\Delta m^2_{31}|$ &$\Delta m^2_{21}$ &$\theta_{23}$&$\theta_{12}$& $\theta_{13}$&$\delta$  &$|\Sigma m_i|$ \\
(TeV) & $(10^{-3}eV^2$)&$(10^{-5}eV^2$)&$(/^0$)&$(/^0$)&$(/^0$)&$(/^0$)&(eV)\\
\hline
2 & 2.46&5.68&44.44 & 32.04 & 8.35& 239.95 &0.114\\ 
4 & 2.44&6.30&44.46&  32.05& 8.35&239.94 &0.112\\ 
6 & 2.42& 6.75& 44.47&32.05&  8.35& 239.93&0.110\\ 
8 &2.41 & 7.05 & 44.48& 32.05 &8.35 &239.93&0.106\\ 
10 &2.40&7.25 &44.48 & 32.06 &8.35&  239.93&0.109\\ 
12 & 2.40& 7.37 &44.49 & 32.06 & 8.35&  239.93&0.108\\ 
14 &2.39&7.49&44.49& 32.06&  8.35&  239.92&0.108\\
\hline 
\end{tabular} 
\caption{\footnotesize{ Effects on the output of neutrino parameters at low energy scale, on varying $m_s$ for IH case ($\tan\beta =40$, $M_R \sim 10^{14}$ GeV).}}
\label{tab:9}
\end{table}

\begin{table}[H]
\begin{tabular}{cccccccc}
\hline  
$m_s$ scale &$|{\Delta m^2_{31}}|$ &$\Delta m^2_{21}$ &$\theta_{23}$&$\theta_{12}$& $\theta_{13}$&$\delta$   &$|\Sigma m_i|$  \\
(TeV) & $(10^{-3}eV^2$)&$(10^{-5}eV^2$)&$(/^0$)&$(/^0$)&$(/^0$)&$(/^0$)&(eV)\\
\hline
2 & 2.37&5.56&44.82 & 32.44 & 8.29&   239.95&0.111\\ 
4 & 2.34&6.11&44.84&  32.45 & 8.29&   239.94 &0.109\\ 
6 & 2.33& 6.55& 44.85&32.45&   8.29&    239.94&0.107\\ 
8 &2.32 & 6.73 & 44.85& 32.45 & 8.29 &   239.93&0.107\\ 
10 &2.31&6.91 &44.86 & 32.46 &8.29 &  239.93&0.106\\ 
12 & 2.30& 7.15 &44.87 & 32.46 & 8.29&   239.93&0.105\\ 
14 &2.30&7.23&44.87& 32.46& 8.29&   239.93&0.105\\
\hline 
\end{tabular} 
\caption{\footnotesize{ Effects on the output of neutrino parameters at low energy scale, on varying $m_s$ for IH case ($\tan\beta =40$, $M_R \sim 10^{15}$ GeV).}}
\label{tab:10}
\end{table}

\begin{table}[H]
\begin{tabular}{cccccccc}
\hline  
$m_s$ scale &$|\Delta m^2_{31}|$ &$\Delta m^2_{21}$ &$\theta_{23}$&$\theta_{12}$& $\theta_{13}$&$\delta$   &$|\Sigma m_i|$  \\
(TeV) & $(10^{-3}eV^2$)&$(10^{-5}eV^2$)&$(/^0$)&$(/^0$)&$(/^0$)&$(/^0$)&(eV)\\
\hline
2 & 2.51&5.22&44.59 & 31.98 & 8.54&  239.94 &0.118\\ 
4 & 2.49&5.73&44.60&  31.98 & 8.54&   239.93 &0.116\\ 
6 & 2.47& 6.14& 44.61&31.99&  8.54&    239.93&0.115\\ 
8 &2.47 & 6.26 & 44.62&  31.99 & 8.54 &   239.92&0.114\\ 
10 &2.46&6.46 &44.62 & 31.99 &8.54 &  239.92&0.114\\ 
12 & 2.45& 6.68 &44.63 & 32.00 & 8.54&   239.92&0.113\\ 
14 &2.45&6.74&44.63& 32.00&8.54&  239.92&0.112\\
\hline 
\end{tabular} 
\caption{\footnotesize{ Effects on the output of neutrino parameters at low energy scale, on varying $m_s$ for IH case ($\tan\beta =40$, $M_R \sim 10^{16}$ GeV). Four different choices of $M_R$ scale are presented.}}
\label{tab:11}
\end{table}

\clearpage

\section{Summary and Discussion}
      To summarize, as a continuation of the earlier investigations  \cite{da} we have studied the running of the RGEs from high scale to low scale and possible effects of the variations of SUSY breaking scale $m_s$ in the range (2-14) TeV on two neutrino mass parameters and three mixing angles, including phases, in both normal \cite{dac} and inverted hierarchical neutrino mass models. Though $\tan{\beta}$ is arbitrary, we consider a fixed value of $\tan{\beta} = 40$ for simplicity in both cases. Effects on the variation of seesaw scale in the range $(10^{13}-10^{16}$) GeV along with the variation of $m_s$, is also studied numerically. Significant effects have been observed for the increase of mass parameters and mixing angles with the increase of SUSY  breaking scale in NH case, but decreases with $m_s$ scale for IH case. The corresponding changes in CP phases are insensitive on variation of $m_s$ scale. The analysis is consistent with the latest cosmological bound on the sum of three neutrino masses.
      
       The present analysis can be applied to check the validity of certain mixing patterns such as Tribimaximal and Golden ratio mixing patterns at high energy scale \cite{gol,gol1}.

\section*{APPENDIX A}
\subsubsection*{RGEs for gauge couplings \cite{rg1}:}
The two loop RGEs for gauge couplings  are given by
\begin{equation}
 \frac{dg_i}{dt}= \frac{b_i}{16 \pi^2} g_i^3  + \frac{1}{(16 \pi^2)^2}\biggl[ \sum_{\mathclap{j=1}}^{3}b_{ij} g_i ^3 g_j ^2 - \sum_{\mathclap{j=t, b, \tau}} a_{ij}g_i ^3 h_j ^2 \biggl],
\end{equation} 

\bigskip
where t = ln $\mu$ and $ b_i ,b_{ij}, a_{ij}$ are $\beta$ function coefficients in MSSM, 
$$ b_{i} =
\left(\begin{array}{ccc}
6.6,  & 1.0,  &   -3.0
\end{array}\right) ,
b_{ij}=
\left(\begin{array}{ccc}
7.96 & 5.40 & 17.60 \\ 
1.80 & 25.00 & 24.00 \\ 
2.20 & 9.00 & 14.00
\end{array}\right),$$
$$ a_{ij}=
\left(\begin{array}{ccc}
5.2 & 2.8 & 3.6 \\ 
6.0 & 6.0 & 2.0 \\ 
4.0 & 4.0 & 0.0
\end{array}\right)$$ 
 and, for non-supersymmetric case, we have\\
 
$$ 
b_i=
\left(\begin{array}{ccc}
4.100,  & -3.167,  &   -7.00
\end{array}\right) ,
g_{ij}=
\left(\begin{array}{ccc}
3.98 & 2.70 & 8.8 \\ 
0.90 & 5.83 & 12.0 \\ 
1.10 & 4.50 & -26.0
\end{array}\right),$$  and 
$$a_{ij}=
\left(\begin{array}{ccc}
0.85 & 0.5 & 0.5 \\
1.50 & 1.5 & 0.5 \\ 
2.00 & 2.0 & 0.0
\end{array}\right).$$  \\

\subsubsection*{Two-loop RGEs for Yukawa couplings and quartic Higgs coupling \cite{rg1}:} 
  For MSSM,
\begin{eqnarray}
\frac{dh_t}{dt} &=&\frac{h_t}{16 \pi^2}\biggl(6h_{t}^2 + h_{b}^2 -\sum_{\mathclap{i=1}}^{3} c_i g_{i}^2 \biggl)
+\frac{h_t}{(16 \pi^2)^2}\biggl[\sum_{\mathclap{i=1}}\biggl(c_i b_i +\frac{c_{i}^2}{2}\biggl)g_{i}^4  + g_{1}^2 g_{2}^2 \biggl. \nonumber\\
& & + \frac{136}{45} g_{1}^2 g_{3}^2 + 8 g_{2}^2 g_{3}^2 +\biggl(\frac{6}{5}g_{1}^2+6 g_{2}^2+16g_{3}^2\biggl)h_{t}^2+ \frac{2}{5}g_{1}^2 h_{b}^2 -22h_{t}^4 \nonumber\\
&& \biggl.- 5 h_{b}^4-5h_{t}^2h_{b}^2 - h_{b}^2 h_{\tau}^2\biggl],
\end{eqnarray}

\begin{eqnarray}
\frac{dh_b}{dt} &=&\frac{h_b}{16 \pi^2}\biggl(6h_{b}^2 + h_{\tau}^{2}+ h_{t}^2- \sum_{\mathclap{i=1}}^{3} c_{i}^{'} g_{i}^2\biggl)+\frac{h_{b}}{(16 \pi^{2})^2}\biggl[ \sum_{\mathclap{i=1}} \biggl(c_{i}^{'} b_i +\frac{c{i}^{'2}}{2}\biggl) g_{i}^4 \biggl. \nonumber\\
&&+ g_{1}^2 g_{2}^2 +\frac{8}{9}g_{1}^2 g_{3}^2  + 8g_{2}^2 g_{3}^2+\biggl(\frac{2}{5} g_{1}^2 +6g_{2}^2+16 g_{3}^2\biggl) h_{b}^2 +\frac{4}{5} g_{1}^2 h_{t}^2 + \frac{6}{5}g_{1}^2 h_{\tau}^2   \nonumber\\
&& \biggl.-22h_{b}^4 - 3 h_{\tau}^4-5h_{t}^4 -5h_{b}^2 h_{t}^2 -3h_{b}^2 h_{\tau}^2\biggl],
\end{eqnarray}

\begin{eqnarray}
\frac{dh_\tau}{dt} &=&\frac{h_\tau}{16 \pi^2}\biggl(4h_{\tau}^2 +3 h_{b}^{2}- \sum_{\mathclap{i=1}}^{3} c_{i}^{''} g_{i}^2\biggl)+\frac{h_{\tau}}{(16 \pi^{2})^2}\biggl[ \sum_{\mathclap{i=1}} \biggl(c_{i}^{''} b_i +\frac{c{i}^{''2}}{2}\biggl)
 g_{i}^4 \biggl. \nonumber\\
&& \biggl.+\frac{9}{5} g_{1}^2 g_{2}^2 + \biggl(\frac{6}{5}g_{1}^2+6 g_{2}^2\biggl) h_{\tau}^2 +\biggl(\frac{-2}{5} g_{1}^2 +16g_{3}^2\biggl)h_{b}^2 +9 g_{b}^4 \biggl. \nonumber\\
&& \biggl.- 10 h_{\tau}^4-3 h_{b}^2 h_{t}^2  -9h_{b}^2 h_{\tau}^2\biggl],
\end{eqnarray}

where  $$c_{i}=
\left(\begin{array}{ccc}
\frac{13}{15}, & 3, & \frac{16}{13}
\end{array}\right) , 
c_{i}^{'}=
\left(\begin{array}{ccc}
\frac{7}{15}, & 3, & \frac{16}{3}  
\end{array}\right)$$  and
$$c_{i}^{''}=
\left(\begin{array}{ccc}
\frac{9}{5} ,& 3 ,& 0  
\end{array}\right).$$  
For non-supersymmetric case,

\begin{eqnarray}
\frac{dh_t}{dt} &=&\frac{h_t}{16 \pi^2}\biggl(\frac{3}{2} h_{t}^2 - \frac{3}{2} h_{b}^2 +Y_{2}(S) - \sum_{\mathclap{i=1}}^{3} c_i g_{i}^2\biggl) + \frac{h_t}{(16 \pi^2)^2}\biggl[\biggl(\frac{1187}{600}\biggl)g_{i}^4 - \frac{23}{4} g_{2}^4  \biggl. \nonumber\\
&&\biggl.- 108 g_{3}^4-\frac{9}{20} g_{1}^2 g_{2}^2+ \frac{19}{15} g_{1}^{2} g_{3}^2 +9 g_{3}^2 g_{2}^2 +\biggl(\frac{223}{80}g_{1}^2+\frac{135}{16} g_{2}^2 +16g_{3}^2\biggl)h_{t}^2  \biggl. \nonumber\\
&&-\biggl( \frac{43}{80}g_{1}^2- \frac{9}{16} g_{2}^2+ 16 g_{3}^2 \biggl) h_{b}^2+\frac{5}{2} Y_{4}(S) - 2 \lambda \biggl(3h_{t}^2+ h_{b}^2\biggl)+ \frac{3}{2} h_{t}^4-\frac{5}{4} h_{t}^2 h_{b}^2   \biggl. \nonumber\\
&&+\frac{11}{4} h_{b}^4+ Y_{2}(S)\biggl(\frac{5}{4}h_{b}^2 - \frac{9}{4}h_{t}^2\biggl)
- \eta _{4}(S)+\frac{3}{2}\lambda^2 \biggl],
\end{eqnarray}

\begin{eqnarray}
\frac{dh_b}{dt} &=&\frac{h_b}{16 \pi^2}\biggl(\frac{3}{2} h_{b}^2 - \frac{3}{2} h_{t}^2 +Y_{2}(S) - \sum_{\mathclap{i=1}}^{3} c_{i}^{'} g_{i}^2 \biggl)+\frac{h_b}{(16 \pi^2)^2}\biggl[ -\frac{127}{600}g_{1}^4-  \frac{23}{4} g_{2}^4 - 108 g_{3}^4 \biggl. \nonumber\\
&& \biggl. - \frac{27}{20} g_{1}^2 g_{2}^2+ \frac{31}{15} g_{1}^{2} g_{3}^2 +9 g_{3}^2 g_{2}^2 -\biggl (\frac{79}{80}g_{1}^2-\frac{9}{16} g_{2}^2 +16g_{3}^2\biggl)h_{t}^2 + \biggl( \frac{187}{80}g_{1}^2+ \frac{135}{16} g_{2}^2  \biggl. \nonumber\\
&&+ 16 g_{3}^2 \biggl) h_{b}^2 +\frac{5}{2} Y_{4}(S)- 2 \lambda \biggl(3h_{t}^2+3 h_{b}^2\biggl) + \frac{3}{2} h_{b}^4-\frac{5}{4} h_{t}^2 h_{b}^2 +\frac{11}{4} h_{t}^4  \biggl. \nonumber\\
 &&\biggl.+ Y_{2}(S)\biggl(\frac{5}{4}h_{t}^2 - \frac{9}{4}h_{b}^2\biggl) - \eta _{4}(S)+\frac{3}{2}\lambda^2 \biggl],
\end{eqnarray}

\begin{eqnarray}
\frac{dh_\tau}{dt} &=&\frac{h_\tau}{16 \pi^2}\biggl(\frac{3}{2} h_{\tau}^2 + Y_{2}(S) - \sum_{\mathclap{i=1}}^{3} c_i^{''}g_{i}^2 \biggl)
+\frac{h_\tau}{(16 \pi^2)^2}\biggl[ \frac{1371}{200}g_{1}^4-  \frac{23}{4} g_{2}^4 -  \frac{27}{20} g_{1}^2 g_{2}^2 \biggl. \nonumber\\
&& \biggl.+\biggl(\frac{387}{80}g_{1}^2+ \frac{135}{16} g_{2}^2 \biggl)h_{\tau}^2 +\frac{5}{2} Y_{4}(S) - 6\lambda h_{t}^2+ \frac{3}{2}h_{\tau}^4 - \frac{9}{4}Y_{2}(S)h_{\tau}^2 \biggl. \nonumber\\
&&\biggl.- \eta _{4}(S) +\frac{3}{2} \lambda^2\biggl],
\end{eqnarray}

\begin{eqnarray}
\frac{d\lambda}{dt} &=&\frac{1}{16 \pi^2}\biggl[\frac{9}{4}\biggl(\frac{3}{25} g_{1}^4 + \frac{2}{5} g_{2}^2 g_{1}^2 + g_{2}^4 \biggl)-\biggl(\frac{9}{5} g_{1}^2 +9 g_{2}^2\biggl)\lambda 
+ 4 Y_{2}(S) \lambda -4H(S) +12 \lambda^2\biggl]
\biggl. \nonumber\\
&&\biggl.+\frac{1}{(16 \pi^2)^2}\biggl[ -78 \lambda^3 + 18\biggl(\frac{3}{5}g_{1}^2+ 3  g_{2}^2 \biggl)\lambda^2 +\biggl(- \frac{73}{8} g_{2}^4 + \frac{117}{20} g_{1}^{2} g_{2}^2 + \frac{1887}{200}g_{1}^4\biggl)\lambda\biggl. \nonumber\\
&&\biggl.+\frac{305}{8} g_{2}^6  - \frac{867}{120}g_{1}^2 g_{2}^4- \frac{1677}{200} g_{1}^4 g_{2}^2-\frac{3411}{1000} g_{1}^6 - 64 g_{3}^2\biggl(h_{t}^4+ h_{b}^4\biggl)- \frac{8}{5} g_{1}^2\biggl(2 h_{t}^4 - h_{b}^4 \biggl. \nonumber\\
&&\biggl.+3 h_{\tau}^4 \biggl)- \frac{3}{2}g_{2}^4 Y_{2}(S)+10 \lambda Y_{4}(S)+ \frac{3}{5}g_{1}^2(- \frac{57}{10}g_{1}^2 + 21 g_{2}^2)h_{t}^2+\biggl(\frac{3}{2}g_{1}^2 +9 g_{2}^2\biggl)h_{b}^2 \biggl. \nonumber\\
&&\biggl. +\biggl(-\frac{15}{2} g_{1}^2 +11 g_{2}^2\biggl)h_{\tau}^2 -24 \lambda^2 Y_{2}(S)-
 \lambda H(S)+6 \lambda h_{t}^2 h_{b}^2 +20 \biggl(3h_{t}^6 +3 h_{b}^6+h_{\tau}^6\biggl)\biggl. \nonumber\\
 &&\biggl. -12\biggl(h_{t}^4 h_{b}^2+h_{t}^2 h_{b}^4\biggl)\biggl],
\end{eqnarray}
where 
 $$ Y_{2}(S)=3 h_{t}^2+ 3 h_{b}^2+ h_{\tau}^2 ,$$
$$ Y_{4}(S)=\frac{1}{3}\biggl[3 \Sigma c_{i} g_{i}^2 h_{t}^2+3 \Sigma c_{i}^{'} g_{i}^2 h_{b}^2+ 3 \Sigma c_{i}^{''} g_{i}^2 h_{\tau}^2\biggl],$$

$$ H(S)=3 h_{t}^4+ 3 h_{b}^4+ h_{\tau}^4, $$

 $$\eta_{4}(S)=\frac{9}{4} \biggl[3 h_{t}^4 + 3 h_{b}^4 + h_{\tau}^4 -\frac{2}{3} h_{t}^2 h_{b}^2\biggl]$$

and $\displaystyle \lambda=  \frac{m_{h}^2}{v_0^2}$ is the Higgs self-coupling, $ m_h$ = 125.78 $\pm $ 0.26 GeV  is the Higgs mass \cite{higgs} and  $v_0$ = 174 GeV is the vacuum expectation value.
\\

The beta function coefficients for  non-SUSY case are given below:

$ \displaystyle c_{i}=
\left(\begin{array}{ccc}
0.85, & 2.25, & 8.00
\end{array}\right), $ 
$\displaystyle c_{i}^{'}=
\left(\begin{array}{ccc}
0.25 & 2.25, & 8.00  
\end{array}\right)$
and
$\displaystyle c_{i}^{''}=
\left(\begin{array}{ccc}
2.25 ,& 2.25 ,& 0.00  
\end{array}\right).$
\section*{APPENDIX B}

\subsubsection*{ RGEs for three neutrino mixing angles and phases \cite{rg2}:(neglecting higher order of $\theta_{13}$ )}

\begin{equation}
\dot{\theta}_{12}= -\frac{Cy_{\tau}^2}{32 \pi^2} \sin2\theta_{12} s_{23}^2 \frac{|m_1 e^{i\psi_1} +m_2 e^{i\psi_2}|^2}{\Delta{m_{21}^2}},
\label{t12}
\end{equation}
$$
\dot{\theta}_{13}= \frac{Cy_{\tau}^2}{32 \pi^2}\sin2\theta_{12}\sin2\theta_{23}\frac{m_3}{\Delta{m_{31}^2}(1+\xi)}  
$$
\begin{equation}
 \times \biggl[ m_1 \cos(\psi_{1}-\delta) -(1+\xi)m_2 \cos(\psi_2 - \delta)- \xi m_3 \cos\delta \biggl],
\end{equation}
\begin{equation}
\dot{\theta}_{23}= -\frac{C y_{\tau}^2}{32 \pi^2} \sin2\theta_{23}\frac{1}{\Delta m_{31}^2}\left[ c_{12}^2 |m_2 e^{i\psi_{2}} +m_3|^2 + s_{12}^2 \frac{|m_1 e^{i\psi_1} +m_3|^2}{1+\xi}\right],
\end{equation}
\label{t23}

 where $\Delta m_{21}^2 = m_2^2 - m_1 ^2$ , $\Delta m_{31}^2 = m_3^2 - m_1 ^2$ and $\xi = \frac{\Delta m_{21}^2 }{\Delta m_{31}^2 }$. 
\subsubsection*{ RGEs for the three phases \cite{rg2}:}
 For  dirac  phase $\delta $ :
\begin{equation}
 \dot{\delta}= \frac{C y_{\tau}^2}{32 \pi^2} \frac{\delta^{(-1)}}{\theta_{13}} + \frac{C y_{\tau}^2}{8 \pi^2} \delta^{(0)}, 
\end{equation}
where
\begin{eqnarray}
\delta^{(-1)} &=& \sin 2 \theta_{12} \sin 2 \theta_{23}\frac{m_3}{\Delta m_{31}^2(1+\xi) } \times \biggl[ m_1 \sin(\psi_1- \delta) 
\biggl. \nonumber\\
&&\biggl.-(1+\xi) m_2 \sin(\psi_2 - \delta) + \xi m_3 \sin\delta\biggl], 
\end{eqnarray}

\begin{eqnarray}
\delta^{(0)} &=& \frac{m_1 m_2 s_{23}^2 \sin(\psi_1 - \psi_2)}{\Delta m_{21}^2} \biggl. \nonumber\\
&&\biggl.  +  m_3 s_{12}^2\biggl[ \frac{m_1 \cos 2\theta_{23} \sin\psi_1}{\Delta m_{31}^2(1+\xi)}+ \frac{m_2 c_{23}^2 \sin(2\delta - \psi_2)}{\Delta m_{31}^2} \biggl]\biggl. \nonumber\\
&&\biggl. +  m_3 c_{12}^2\biggl[ \frac{m_1 c_{23}^2 \sin(2 \delta - \psi_1)}{\Delta m_{31}^2(1+\xi)}+ \frac{m_2 \cos(2\theta_{23})\sin\psi_2}{\Delta m_{31}^2} \biggl].
\end{eqnarray}

For Majorana phase $\psi_1$ \cite{rg2}:
\begin{eqnarray}
\dot{\psi_1}&=&\frac{C y_{\tau}^2}{8 \pi^2} \left[ m_3 \cos 2 \theta_{23} \frac{m_1 s_{12}^2 \sin \psi_1 + (1+ \xi) m_2 c_{12}^2 \sin \psi_2}{\Delta m_{31}^2(1+ \xi)}\right]\biggl. \nonumber\\
 &&\biggl. + \frac{C y_{\tau}^2}{8 \pi^2}\left[ \frac{m_1 m_2 c_{12}^2 s_{23}^2 sin(\psi_1 - \psi_2)}{
\Delta m_{21}^2}\right] 
\end{eqnarray}
 For Majorana phase $\psi_2$:
\begin{eqnarray}
 \dot{\psi_2}&=&\frac{C y_{\tau}^2}{8 \pi^2}\left[  m_3 \cos 2 \theta_{23} \frac{m_1 s_{12}^2 \sin \psi_1 +(1+\xi) m_2 c_{12}^2 \sin \psi_2  }{\Delta m_{31}^2 (1+\xi)}\right]\biggl. \nonumber\\
 &&\biggl.+\frac{C y_{\tau}^2}{8 \pi^2}\left[ \frac{m_1 m_2 s_{12}^2 s_{23}^2 \sin(\psi_1 - \psi_2
)}{\Delta m_{21}^2} \right] 
\end{eqnarray}
\subsubsection*{RGEs for neutrino mass eigenvalues \cite{rg2}:}
\begin{equation}
\dot{m_1} = \frac{1}{16 \pi^2}\left[  \alpha + C y_{\tau}^2 (2 s_{12}^2 s_{23}^2 +F_1)\right]m_1,
\end{equation}
\begin{equation}
\dot{m_2} = \frac{1}{16 \pi^2}\left[  \alpha + C y_{\tau}^2 (2 c_{12}^2 s_{23}^2 +F_2)\right]m_2,
\end{equation}
\begin{equation}
\dot{m_3} = \frac{1}{16 \pi^2}\left[  \alpha + 2C y_{\tau}^2 c_{13}^2 c_{23} \right]m_3,
\end{equation}
where
\begin{equation}
 F_1 = - s_{13}\sin 2 \theta_{12} \sin 2 \theta_{23} \cos \delta + 2 s_{13}^2 c_{12}^2 c_{23}^2,
 \end{equation}
 \begin{equation}
 F_2 =  s_{13}\sin 2 \theta_{12} \sin 2 \theta_{23} \cos \delta + 2 s_{13}^2 s_{12}^2 s_{23}^2.
 \end{equation}
  
For MSSM case:
  
$$\alpha= -\frac{6}{5} g_{1}^2 - 6 g_2 ^2 +6 y_t^2 $$ and $$C=1.$$

For SM case: 
 
$$\alpha= -3 g_{2}^2 + 2 y_{\tau}^2  +6 y_{t}^2 +6 y_{b}^2 + \lambda,$$
$$C=-\frac{3}{2}$$ and $\lambda$ is the Higgs self-coupling in the SM.
\section*{Acknowledgements}
One of the author (KHD) would like to thank Manipur University for  financial support. 
\bibliographystyle{ieeetr}
\bibliography{ref}

\begin{thebibliography}{10}

\bibitem{a1}
D.~Adey {\em et~al.}, ``{Measurement of the Electron Antineutrino Oscillation
  with 1958 Days of Operation at Daya Bay},'' {\em Phys. Rev. Lett.}, vol.~121,
  no.~24, p.~241805, 2018.

\bibitem{a2}
Y.~Abe {\em et~al.}, ``{Indication of Reactor $\bar{\nu}_e$ Disappearance in
  the Double Chooz Experiment},'' {\em Phys. Rev. Lett.}, vol.~108, p.~131801,
  2012.

\bibitem{a3}
G.~Bak {\em et~al.}, ``{Measurement of Reactor Antineutrino Oscillation
  Amplitude and Frequency at RENO},'' {\em Phys. Rev. Lett.}, vol.~121, no.~20,
  p.~201801, 2018.

\bibitem{b1}
J.~L. Miller, ``{Accelerator experiments are closing in on neutrino CP
  violation},'' {\em Phys. Today}, vol.~73, no.~6, pp.~14--16, 2020.

\bibitem{b2}
K.~Abe {\em et~al.}, ``{Constraint on the matter\textendash{}antimatter
  symmetry-violating phase in neutrino oscillations},'' {\em Nature}, vol.~580,
  no.~7803, pp.~339--344, 2020.
\newblock [Erratum: Nature 583, E16 (2020)].

\bibitem{c}
Y.~Fukuda {\em et~al.}, ``{Evidence for oscillation of atmospheric
  neutrinos},'' {\em Phys. Rev. Lett.}, vol.~81, pp.~1562--1567, 1998.

\bibitem{c1}
Q.~R. Ahmad {\em et~al.}, ``{Direct evidence for neutrino flavor transformation
  from neutral current interactions in the Sudbury Neutrino Observatory},''
  {\em Phys. Rev. Lett.}, vol.~89, p.~011301, 2002.

\bibitem{c2}
M.~C. Gonzalez-Garcia, M.~Maltoni, and T.~Schwetz, ``{NuFIT: Three-Flavour
  Global Analyses of Neutrino Oscillation Experiments},'' {\em Universe},
  vol.~7, no.~12, p.~459, 2021.

\bibitem{g}
Alam {\em et~al.}, ``Completed sdss-iv extended baryon oscillation
  spectroscopic survey: Cosmological implications from two decades of
  spectroscopic surveys at the apache point observatory,'' {\em Phys. Rev. D},
  vol.~103, p.~083533, Apr 2021.

\bibitem{h}
N.~Aghanim {\em et~al.}, ``{Planck 2018 results. VI. Cosmological
  parameters},'' {\em Astron. Astrophys.}, vol.~641, p.~A6, 2020.
\newblock [Erratum: Astron.Astrophys. 652, C4 (2021)].

\bibitem{j2}
M.~Agostini {\em et~al.}, ``{Final Results of GERDA on the Search for
  Neutrinoless Double-$\beta$ Decay},'' {\em Phys. Rev. Lett.}, vol.~125,
  no.~25, p.~252502, 2020.

\bibitem{j1}
D.~Q. Adams {\em et~al.}, ``{Improved Limit on Neutrinoless Double-Beta Decay
  in $^{130}$Te with CUORE},'' {\em Phys. Rev. Lett.}, vol.~124, no.~12,
  p.~122501, 2020.

\bibitem{k}
M.~Aker {\em et~al.}, ``{Direct neutrino-mass measurement with sub-electronvolt
  sensitivity},'' {\em Nature Phys.}, vol.~18, no.~2, pp.~160--166, 2022.

\bibitem{s1}
S.~Dimopoulos, S.~Raby, and F.~Wilczek, ``{Supersymmetry and the Scale of
  Unification},'' {\em Phys. Rev. D}, vol.~24, pp.~1681--1683, 1981.

\bibitem{s2}
N.~Haba and T.~Ota, ``{Vanishing dimension five proton decay operators in the
  SU(5) SUSY GUT},'' {\em Acta Phys. Polon. B}, vol.~39, pp.~1901--1912, 2008.

\bibitem{s3}
S.~Heinemeyer, Mondragon, {\em et~al.}, ``{The Higgs boson discovery: recent
  implications for the Finite Unified Theories and SUSY breaking scale},'' {\em
  PoS}, vol.~CORFU2017, p.~081, 2018.

\bibitem{u1}
Y.~Yamada, ``{SUSY and GUT threshold effects in SUSY SU(5) models},'' {\em Z.
  Phys. C}, vol.~60, pp.~83--94, 1993.

\bibitem{u2}
N.~N. Singh and S.~B. Singh, ``{Third generation Yukawa couplings unification
  in supersymmetric SO(10) model},'' {\em Eur. Phys. J. C}, vol.~5,
  pp.~363--367, 1998.

\bibitem{da}
K.~S. Singh and N.~N. Singh, ``{Effects of the Variation of SUSY Breaking Scale
  on Yukawa and Gauge Couplings Unification},'' {\em Adv. High Energy Phys.},
  vol.~2015, p.~652029, 2015.

\bibitem{ew1}
S.~Dawson, ``{Introduction to Electroweak Symmetry Breaking},'' {\em AIP Conf.
  Proc.}, vol.~1116, no.~1, pp.~11--34, 2009.

\bibitem{ms2}
S.~Dawson, ``{The MSSM and why it works},'' {\em Theoretical Advanced Study
  Institute in Elementary Particle Physics ,TASI'97. Proc.}, pp.~261--339,
  1997.

\bibitem{lhc1}
R.~Goncalo, S.~Guindon, and V.~Jain, ``{Sensitivity of LHC experiments to the
  $t\bar{t}H$ final state, with $H \rightarrow b\bar{b}$, at center of mass
  energy of 14 TeV},'' {\em Contribution to Community Summer Study 2013 on the
  Future of U.S. Particle Physics: Snowmass on the Mississippi ,CSS2013.},
  p.~4, 2013.

\bibitem{lhc2}
A.~K. Chaudhuri, ``{Large elliptic flow in low multiplicity pp collisions at
  LHC energy s**(1/2) = 14-TeV},'' {\em Phys. Lett. B}, vol.~692, pp.~15--19,
  2010.

\bibitem{ss1}
L.~Randall and M.~Reece, ``{Single-Scale Natural SUSY},'' {\em JHEP}, vol.~08,
  p.~088, 2013.

\bibitem{ss2}
K.~Yonekura, ``{Single scale model of SUSY breaking, gauge mediation, and dark
  matter},'' {\em Soryushiron Kenkyu Electron.}, vol.~119, p.~1, 2011.

\bibitem{st1}
R.~L. Arnowitt and P.~Nath, ``{SUSY mass spectrum in SU(5) supergravity grand
  unification},'' {\em Phys. Rev. Lett.}, vol.~69, pp.~725--728, 1992.

\bibitem{st2}
Y.~Yamada, ``{SUSY and GUT threshold effects in SUSY SU(5) models},'' {\em Z.
  Phys. C}, vol.~60, pp.~83--94, 1993.

\bibitem{rg2}
S.~Antusch, J.~Kersten, M.~Lindner, and M.~Ratz, ``{Running neutrino masses,
  mixings and CP phases: Analytical results and phenomenological
  consequences},'' {\em Nucl. Phys. B}, vol.~674, pp.~401--433, 2003.

\bibitem{beta}
D.~R.~T. Jones and L.~Mezincescu, ``{The Beta Function in Supersymmetric
  {Yang-Mills} Theory},'' {\em Phys. Lett. B}, vol.~136, pp.~242--244, 1984.

\bibitem{dac}
K.~S. Singh, S.~Roy, and N.~N. Singh, ``{Stability of neutrino parameters and
  self-complementarity relation with varying SUSY breaking scale},'' {\em Phys.
  Rev. D}, vol.~97, no.~5, p.~055038, 2018.

\bibitem{rg1}
V.~D. Barger, M.~S. Berger, and P.~Ohmann, ``{Supersymmetric grand unified
  theories: Two loop evolution of gauge and Yukawa couplings},'' {\em Phys.
  Rev. D}, vol.~47, pp.~1093--1113, 1993.

\bibitem{d}
P.~Zyla {\em et~al.}, ``{Review of Particle Physics},'' {\em PTEP}, vol.~2020,
  no.~8, p.~2093, 2020.
\newblock and 2021 update.

\bibitem{bb1}
J.~E. Bjorkman and D.~R.~T. Jones, ``{The Unification Mass, $Sin^2\theta_W$ and
  $M_b / M_\tau$ in Nonminimal Supersymmetric SU(5)},'' {\em Nucl. Phys. B},
  vol.~259, p.~533, 1985.

\bibitem{qcd}
M.~Patgiri and N.~N. Singh, ``{New uncertainties in QCD-QED rescaling factors
  using quadrature method},'' {\em Pramana}, vol.~65, pp.~1015--1025, 2006.

\bibitem{gol}
Y.~Kajiyama, M.~Raidal, and A.~Strumia, ``{The Golden ratio prediction for the
  solar neutrino mixing},'' {\em Phys. Rev. D}, vol.~76, p.~117301, 2007.

\bibitem{gol1}
L.~L. Everett and A.~J. Stuart, ``{Icosahedral (A(5)) Family Symmetry and the
  Golden Ratio Prediction for Solar Neutrino Mixing},'' {\em Phys. Rev. D},
  vol.~79, p.~085005, 2009.

\bibitem{higgs}
A.~Sirunyan, A.~Tumasyan, and other, ``A measurement of the higgs boson mass in
  the diphoton decay channel,'' {\em Physics Letters B}, vol.~805, p.~135425,
  2020.

\end{thebibliography}
\end{document}